\documentclass[10pt,journal,twoside]{IEEEtran}
\usepackage[mathscr]{eucal}
\usepackage{amssymb}
\usepackage{amsthm}

\usepackage{multirow}
\usepackage{inputenc}
\usepackage{enumerate}
\usepackage{enumitem}
\usepackage{textcomp}
\usepackage{listings}
\usepackage{array}
\usepackage[switch,mathlines]{lineno}
\usepackage{soul}
\usepackage{xfrac}
\usepackage{graphicx}
\usepackage{cite}
\usepackage[cmex10]{amsmath}
\usepackage{xcolor}
\usepackage{colortbl}
\usepackage{cancel}
\usepackage{siunitx}
\usepackage{nomencl}
\usepackage{multirow}
\usepackage{svg}
\usepackage{doi}
\usepackage{hyperref}
\usepackage{amsmath}

\usepackage{verbatim}
\usepackage{bbm}
\usepackage{url}
\usepackage[caption=false]{subfig}
\usepackage{float}

\usepackage[ruled,norelsize,vlined,linesnumbered]{algorithm2e}
\makeatletter
\newcommand{\removelatexerror}{\let\@latex@error\@gobble}
\makeatother

\usepackage{graphicx}
\usepackage{longtable}
\newlist{steps}{enumerate}{1}
\setlist[steps, 1]{label = Step \arabic*:}
\newtheorem{theorem}{Theorem}

\makeatletter
\def\footnoterule{\kern-3\p@
  \hrule \@width 3.3in \kern 2.6\p@} 
\makeatother

\usepackage{algpseudocode}

\usepackage{stackengine}

\usepackage{algpseudocode}
\usepackage{CJK}
\usepackage[cmex10]{amsmath}
\usepackage{bm}
\usepackage{color}
\usepackage{amsmath,amsthm,amssymb,amsfonts}
\usepackage{flushend}

\allowdisplaybreaks[3]

\newtheorem{lemma}{Lemma}
\newtheorem{remark}{Remark}

\usepackage{arydshln} 
\usepackage{hyperref}
\hypersetup{
     colorlinks   = true,
     linkcolor    = blue,
     citecolor    = blue,
     urlcolor     = blue
}
\makeatletter
\newcommand*{\transpose}{%
  {\mathpalette\@transpose{}}%
}
\newcommand*{\@transpose}[2]{%
  \raisebox{\depth}{$\m@th#1\intercal$}%
}
\makeatother

\usepackage{soul}
\usepackage{tikz}
\usepackage{booktabs}
\usepackage{tabularx}
\usepackage{setspace}
\usepackage{amsmath,amsfonts}
\usepackage{diagbox}
\DeclareMathOperator*{\argmin}{arg\,min}

\ifCLASSINFOpdf
\else
\fi
\renewcommand\nomgroup[1]
{
  \ifthenelse{\equal{#1}{A}}{\item[\textit{A. Abbreviations}]}{%
  \ifthenelse{\equal{#1}{S}}{\item[\textit{B. Sets}]}{
  \ifthenelse{\equal{#1}{V}}{\item[\textit{C. Parameters and Variables}]} \\
  }
  }
  }

\begin{document}
\renewcommand{\ttdefault}{cmtt}
\bstctlcite{IEEEexample:BSTcontrol}
\SetKwComment{Comment}{/* }{ */}
\renewcommand\qedsymbol{$\blacksquare$}

\title{On the Data-Driven Modeling of Price-Responsive Flexible Loads: Formulation and Algorithm}

\author{
{
Mingji Chen,~\IEEEmembership{Student Member, IEEE},
Shuai Lu,
Wei Gu,~\IEEEmembership{Senior Member, IEEE}, \\
Zhaoyang Dong,~\IEEEmembership{Fellow, IEEE},
Yijun Xu,~\IEEEmembership{Senior Member, IEEE},
Jiayi Ding
}

}

\markboth{S\MakeLowercase{ubmitted to} IEEE Transactions on Power Systems}
{Chen \MakeLowercase{\textit{\textit{et al.}}}: On the Data-Driven Modeling of Price-Responsive Flexible Loads: Formulation and Algorithm}
\maketitle
\vspace{-1cm}
\begin{abstract}
The flexible loads in power systems, such as interruptible and transferable loads, are critical flexibility resources for mitigating power imbalances. 
Despite their potential, accurate modeling of these loads is a challenging work and has not received enough attention, limiting their integration into operational frameworks. To bridge this gap, this paper develops a data-driven identification theory and algorithm for price-responsive flexible loads (PRFLs).
First, we introduce PRFL models that capture both static and dynamic decision mechanisms governing their response to electricity price variations.
Second, We develop a data-driven identification framework that explicitly incorporates forecast and measurement errors. Particularly, we give a theoretical analysis to quantify the statistical impact of such noise on parameter estimation.
Third, leveraging the bilevel structure of the identification problem, we propose a Bayesian optimization-based algorithm that features the scalability to large sample sizes and the ability to offer posterior differentiability certificates as byproducts.
Numerical tests demonstrate the effectiveness and superiority of the proposed approach.
\end{abstract}


\vspace{-0.4cm}
\section{Introduction}
\IEEEPARstart{W}{i}th increasing incorporation of renewable energy sources (RES) into the power system\cite{Yu_2024_water}, the inherent uncertainty and intermittent of RES cause significant power imbalances between supply and demand side. In this context, the integration of demand-side flexible loads (FLs), including thermostatically controlled loads\cite{Mathieu2015Arbitraging}, plug-in electric vehicles\cite{Xu2015Hierarchical}, and energy storages\cite{Li2020}, into power systems is becoming a promising solution to address power imbalance \cite{ Callaway2011Achievinga, Mathieu2013State}. 


To incentivize the FLs to offer flexibility for power system operation, there are two main market mechanisms: incentive-based mechanisms that utilize financial subsidies for the adjustments of FLs, and  price-responsive mechanisms that use price signals to motivate the FLs to adjust their response power. The FLs under price-responsive mechanisms (PRFLs) can ensure market fairness, minimize intervention, and facilitate efficient settlement, and thus has been widely adopted by power system operators (PSOs). In this case, it is a fundamental prerequisite for PSOs to develop accurate PRFL models to ensure safe and economical operation\cite{Chicco2020Flexibility}. Given the inherent spatial dispersion and behavioral heterogeneity of individual PRFLs,  a practical scheme to manage PRFLs is to aggregate them in certain area, which requires us to build the aggregate model (AGM) of PRFLs  \cite{Xu2018,Sharma2015Smart}. In current research, two predominant approaches are employed to derive the AGM, including the physics-based ones and the data-driven ones. 
The physics-based approaches adopt a bottom-up construction, initiating from individual PRFL component models to derive AGM through some mathematical approaches, including: (1) the geometric projection techniques that frame this issue as a feasible region projection challenge, exemplified by Minkowski summation \cite{Chicco2020Flexibility} and Fourier-Motzkin elimination \cite{Liu2021,Yikui2021,Patig2022}; (2) the optimization-driven approximations that formulate convex outer \cite{Muller2019Aggregation} or inner \cite{Wen2024} bounds to encapsulate operational constraints; and (3) the heuristic simplification strategies such as non-iterative vertex enumeration \cite{Tan2023Noniterative} and recursive feasible region expansion \cite{Tan2024II,Wen2023Aggregate}.
While these methods provide physical mechanism models, they exhibit two shortcomings: high computational complexity and relying on precise PRFL information. Moreover, because of the privacy concerns, the PRFLs are not necessarily willing to  share their detailed information with PSOs,  hindering the practical feasibility of the above methods.

Alternatively, the data-driven approaches employ a top-down design philosophy, extracting price-response mechanisms from historical operational data. The early research relies on shallow learning architectures like linear regression and artificial neural networks \cite{Govender2019},  suffering from data quality dependencies and weak physical interpretability. Some subsequent research integrated physical priors into the data-driven frameworks. For example, Taheri \textit{et al.} \cite{Taheri2022}  proposed to use the convex quadratic classifiers that embedded thermodynamic constraints to  enhance the interpretability. However, the inherent restrictive constraint formulations and device parameter requirements limit the application of this method.  Consequently, the inverse optimization (IO) based method \cite{Awadalla2024,FernndezBlanco2021InverseOW,Bian_2024} is proposed for the PRFL modeling, which overcomes the above limitations by embedding different prior physics structures of PRFL in data-driven frameworks.  

Resorting to IO framework, Tan \textit{et al}. \cite{Tan2023Datadriven} identified the upper and lower bound parameters in the AGM model of the PRFLs by a novel Newton-based algorithm. Lyu \textit{et al}. \cite{Lyu2024} proposed a co-regularized IO architecture to reduce computation time and enhance the robustness against data noise. However, some critical parameters such as storage efficiency are still unable to be identified by these methods, and these identification frameworks can hardly extend to other response mechanisms, such as the dynamic mechanisms.  In the analysis of this IO problem solution, Lu \textit{et al}.  \cite{Lu2024Solution} first underscores the uniqueness of model solutions and emphasizes proper prior model selection, while demonstrating conventional KKT-based methods' limited scalability with sample size and highlighting the need for noise impact analysis on parameter identifiability.

While the above research has carried out pioneering research in the data-driven modeling of PRFL, there are still some unsolved problems, summarized as follows:
(1) In the modeling, the current identification models usually assume static response mechanisms, failing  in  accounting for the dynamic response mechanisms that may be also implemented in reality. Moreover, the robustness of these models against measurement noise and data perturbations remains inadequately addressed in existing research.
(2) The inherent bilevel structure and non-convex nature of the identification model seriously compromises the effectiveness of conventional algorithms in terms of guaranteeing global optimality while maintaining computational tractability.
(3) Furthermore, as revealed by \cite{Lu2024Solution}, the identification model exhibits solution multiplicity, necessitating developing a dedicated computational framework that can evaluate the performance of the solution a posterior.

To bridge the above-mentioned gaps, we first develop the response models of the PRFLs under static and dynamic decision mechanisms. We then formulate the parameter identification of PRFLs as an IO model and propose an improved Bayesian optimization (BayOpt) based solution method. The main contributions are summarized as follows:
\begin {enumerate}
\item {We develop a novel data-driven identification model for the PRFLs. The proposed identification model is formulated as an inverse optimization problem embedding the forward decision problem of the PRFLs, in which the data noise including the forecast and measurement errors is integrated. Particularly, we theoretically analyze the impact of data noise on the identification results.}
\item {We model both the static and dynamic decision mechanisms of the PRFLs in the forward problem. Also, the forward problem model is further extended to integrate the inner cost of the PRFLs to make it more practical.}
\item {We propose a BayOpt based algorithm for the identification of the PRFLs. The method is scalable to large sample size and inherently provides posterior identifiability certificates as a byproduct. Besides, we introduce a block Cholesky decomposition (BCD) technique, significantly accelerating the inversion of covariance matrices while preserving numerical stability.}

\end {enumerate}
\section{Problem Descriptions}
\label{section2}
This section develops the PRFLs identification framework through physical model derivation and data-driven inverse formulation. The notations are defined as follows: $\mathbf{T}$  denote the index set of the planning horizon; $t \in$\textbf{$\mathbf{T}$} denotes the index of time period;  $T$ and $\Delta t$  indicate the cardinality of  $\mathbf{T}$ and temporal resolution, respectively; 
$\mathbf{N}$ denotes the index set of samples; $n\in \mathbf{N}$ denotes the index of samples;  
$\mathbf{I}_{(\cdot)}$ denotes the index set of the fixed power ($\mathbf{I}_{fix}$)/adjustable power ($\mathbf{I}_{adj}$)/energy storage ($\mathbf{I}_{str}$) components in the PRFLs; $i,j\in \mathbf{I}_{(\cdot)}$ denotes the index of the components;
$p_{(\cdot)}^{i,t}\in \mathbb{R}$ and $e_{(\cdot)}^{i,t}\in \mathbb{R}$ denote instantaneous power/energy scalars, with their boldface counterparts  $P_{(\cdot)}^i \in \mathbb{R}^T$ and $E_{(\cdot)}^i  \in \mathbb{R}^T$ representing the chronological vectors.
For instance, $p_{fix}^{i,t}$ denotes the fixed load power of the $i$-th component at period $t$, and the temporal sequence becomes $P_{fix}^i=[p_{fix}^{i,1},p_{fix}^{i,2},\cdots,p_{fix}^{i,T}]^\top$.
Besides, we use $\hat{(\cdot)}$, $\tilde{(\cdot)}$, and $(\cdot)^*$ to denote the estimated, observed, and true value of the variables, respectively.

\vspace{-0.4cm}
\subsection{Physical Model of PRFL}
\subsubsection{Components of PRFLs}
In this work, we use three components to model the PRFLs, including the fixed load component, the adjustable power component,  and energy storage.  

\textit{Fixed load component $P_{fix} \in \mathbb{R}^T$:} It characterizes price-insensitive, periodically stable demands essential for maintaining operational continuity, encompassing process-critical industrial loads and continuous-operation lighting systems. This component exhibits predictability due to its inherent cyclicity and invariability. 

\textit{Adjustable power component $P_{adj}\in \mathbb{R}^T$:} It characterizes the component in the PRFLs that has time-independent power bounds and the power variation limitation, such as the distributed generators and interruptible loads. The operational characteristics are captured by: 
\begin{subequations}
\begin{equation}
    \underline{p}_{\,adj}^j \leq p_{adj}^{j,t} \leq \overline{p}_{adj}^j \quad \forall t \in \mathbf{T},  j \in \mathbf{I}_{adj}, 
\end{equation}
\begin{equation}
    \underline{r}^j \Delta t \leq p_{adj}^{j,t} - p_{adj}^{j,t - 1} \leq \overline{r}^j \Delta t \quad \forall  t \in \mathbf{T}, j \in \mathbf{I}_{adj},
\end{equation}
\end{subequations}
wherein $\overline{p}_{adj}^j, \underline{p}_{\,adj}^j  \in \mathbb{R}$ represent the upper and lower power bounds,   and $\overline{r}^j, \underline{r}^j \in \mathbb{R}$ denote the ramp-rate limits.

Here, we give the matrix-form model of the adjustable power component.
\begin{subequations}
\begin{equation}
\underline{P}_{adj}^{j} \leq P_{adj}^{j} \leq \overline{P}_{adj}^{j},
\end{equation}
\begin{equation}
\underline{R}^{j} \Delta t \leq M P_{adj}^{j} \leq \overline{R}^{j} \Delta t, \ \forall j \in \mathbf{I}_{adj},
\end{equation}
\end{subequations}
wherein $M \in \mathbb{R}^{T\times T}$ represents the temporal difference matrix with elements  $[ M]_{t,t}=-1$ , $[M]_{t,t+1}=1$ and the rest are 0. $\overline{R}^j=[\overline{r}^{j},\overline{r}^{j},\cdots,\overline{r}^{j}]^\top\in\mathbb{R}^T$ and $\underline{R}^j=[\underline{r}^{j},\underline{r}^{j},\cdots,\underline{r}^{j}]^\top\in\mathbb{R}^T$.

\textit{Energy storage component $P_{str}\in \mathbb{R}^T$:} It characterizes the energy storage systems (EST) with bidirectional power exchange capability, enabling strategic energy arbitrage during price-fluctuating periods. We use the following model to describe this component, as:
\begin{subequations}
\label{str_scalarmodel}
\begin{equation}
\label{p_vb_bound}
\underline{p}_{str}^{i}\leq {p}_{str}^{i,t}\leq \overline{p}_{str}^{i}\quad \forall t\in \mathbf{T},i\in \mathbf{I}_{str},
\end{equation}
\begin{equation}
\label{e_vb_bound}
\underline{e}_{str}^{i}\leq {e}_{str}^{i,t}\leq \overline{e}_{str}^{i}
\quad \forall t\in \mathbf{T},i\in \mathbf{I}_{str},
\end{equation}
\begin{equation}
\label{e_vb_eq}
e_{str}^{i,t}=\sigma^i e_{str}^{i,t-1}+p_{str}^{i,t} \Delta t\quad \forall t\in \mathbf{T},i\in \mathbf{I}_{str},
\end{equation}
wherein $\sigma^i\in(0,1]$ represents the storage efficiency. $\overline{e}_{str}^{i}, \underline{e}_{str}^{i} \in \mathbb{R}$ denote the upper and lower energy bounds. $\overline{p}_{str}^{i},\underline{p}_{str}^{i} \in \mathbb{R}$ describes the charge/discharge upper and lower power limits. If the EST consume power to increase energy storage, then $p_{str}^{i,t} \geq 0$ and vice versa. The equation \eqref{e_vb_eq} can be also rewritten as:
\begin{equation}
\label{e_vb_eq_e0}
{e}_{str}^{i,t}= \sum_{k=1}^{t}(\sigma^{i})^{t-k}p_{str}^{i,k}\Delta t +(\sigma^{i})^t {e}_{str}^{i,0}
\quad \forall t\in \mathbf{T},i\in \mathbf{I}_{str},
\end{equation}
\end{subequations}
wherein ${e}_{str}^{i,0} \in \mathbb{R}$ denotes the initial energy storage of the $i$-th energy storage component.
Further, we can reformulate the model \eqref{str_scalarmodel} in vector space form, as:
\begin{subequations}
\label{p_vb_T_total}
\begin{equation}
\label{p_vb_T}
\underline{P}_{str}^{i} \leq P_{str}^{i} \leq \overline{P}_{str}^{i}\quad \forall i \in \mathbf{I}_{str},
\end{equation}
\begin{equation}
\label{E_vb_T}
\underline{E}_{str}^{i} \leq E_{str}^{i} \leq \overline{E}_{str}^{i}\quad \forall i \in \mathbf{I}_{str},
\end{equation}
\begin{equation}
\label{E_vb_T_eq}
E_{str}^{i} =(\gamma^i)^\top [E_{str}^{i}]_0 + \varUpsilon^i P_{str}^{i}\Delta t\quad \forall i \in \mathbf{I}_{str},
\end{equation}
wherein 
\begin{equation}
\begin{aligned}
    \underline{P}_{str}^{i}=[\underline{p}_{str}^i,\underline{p}_{str}^i,\cdots,\underline{p}_{str}^i]^\top, \ 
    \overline{P}_{str}^{i}=[\overline{p}_{str}^i,\overline{p}_{str}^i,\cdots,\overline{p}_{str}^i]^\top, \\ \notag
    \underline{E}_{str}^{i}=[\underline{e}_{str}^i,\underline{e}_{str}^i,\cdots,\underline{e}_{str}^i]^\top, \ 
    \overline{E}_{str}^{i}=[\overline{e}_{str}^i,\overline{e}_{str}^i,\cdots,\overline{e}_{str}^i]^\top,\\ \notag
\end{aligned}
\end{equation}
and
\begin{equation}
\begin{aligned}
    \gamma^i=[\sigma^i,(\sigma^i)^2,\cdots,(\sigma^i)^T]^\top,\\ \notag
\varUpsilon^i=\begin{bmatrix}
1 & 0 & \cdots  & 0 \\
(\sigma^i)^{2-1} & 1 & \cdots  & 0 \\
\vdots  & \vdots  & \ddots  & \vdots \\
(\sigma^i)^{T-1} & (\sigma^i)^{T-2} & \cdots  & 1
\end{bmatrix}.\notag
\end{aligned}
\end{equation}
\end{subequations}

\subsubsection{Aggregate model of PRFLs}
The aggregate model of the PRFL can be derived from above component models, defined as:
\begin{equation}
\label{eq_sum_model_T}
\mathbf{\Omega}_{T} := \left\{ 
\begin{aligned}
&P_{agg} = {P}_{fix}^{i} +  \textstyle \sum \limits_{j \in \mathbf{I}_{adj}} P_{adj}^j + \textstyle \sum \limits_{i \in \mathbf{I}_{str}} P_{str}^i, \\
& \underline{P}_{adj}^{j} \leq P_{adj}^{j} \leq \overline{P}_{adj}^{j},\\
&\underline{R}^{j} \Delta t \leq M P_{adj}^{j} \leq \overline{R}^{j} \Delta t, \ \forall j \in \mathbf{I}_{adj}\\
& \underline{P}_{str}^{i} \leq P_{str}^{i} \leq \overline{P}_{str}^{i}, \ \underline{E}_{str}^{i} \leq E_{str}^{i} \leq \overline{E}_{str}^{i}, \\
& E_{str}^{i} =(\gamma^i)^\top [E_{str}^{i}]_0 + \varUpsilon^i P_{str}^{i} \Delta t, \ \forall i \in \mathbf{I}_{str},
\end{aligned}
\right \}\in \mathbb{R}^T,
\end{equation}

\vspace{-0.3cm}
\subsection{Identification Problem of PRFLs}
We collect the parameters in the model $\mathbf{\Omega}_T$ into the set $\theta$, as follows:
\begin{equation}
\begin{aligned}
\theta=\{&\underline{p}_{str}^i,\overline{p}_{str}^i,\underline{e}_{str}^i,\overline{e}_{str}^i,{e}_{str}^{i,0},\sigma^i,\\
&\underline{p}^j_{adj},\overline{p}^j_{adj},\underline{r}^j,\overline{r}^j \quad \forall i \in \mathbf{I}_{str},\forall j \in \mathbf{I}_{adj} \}.
\end{aligned}
\end{equation}

\begin{figure}[tb]
\centering
\footnotesize
\includegraphics[width=1\linewidth]{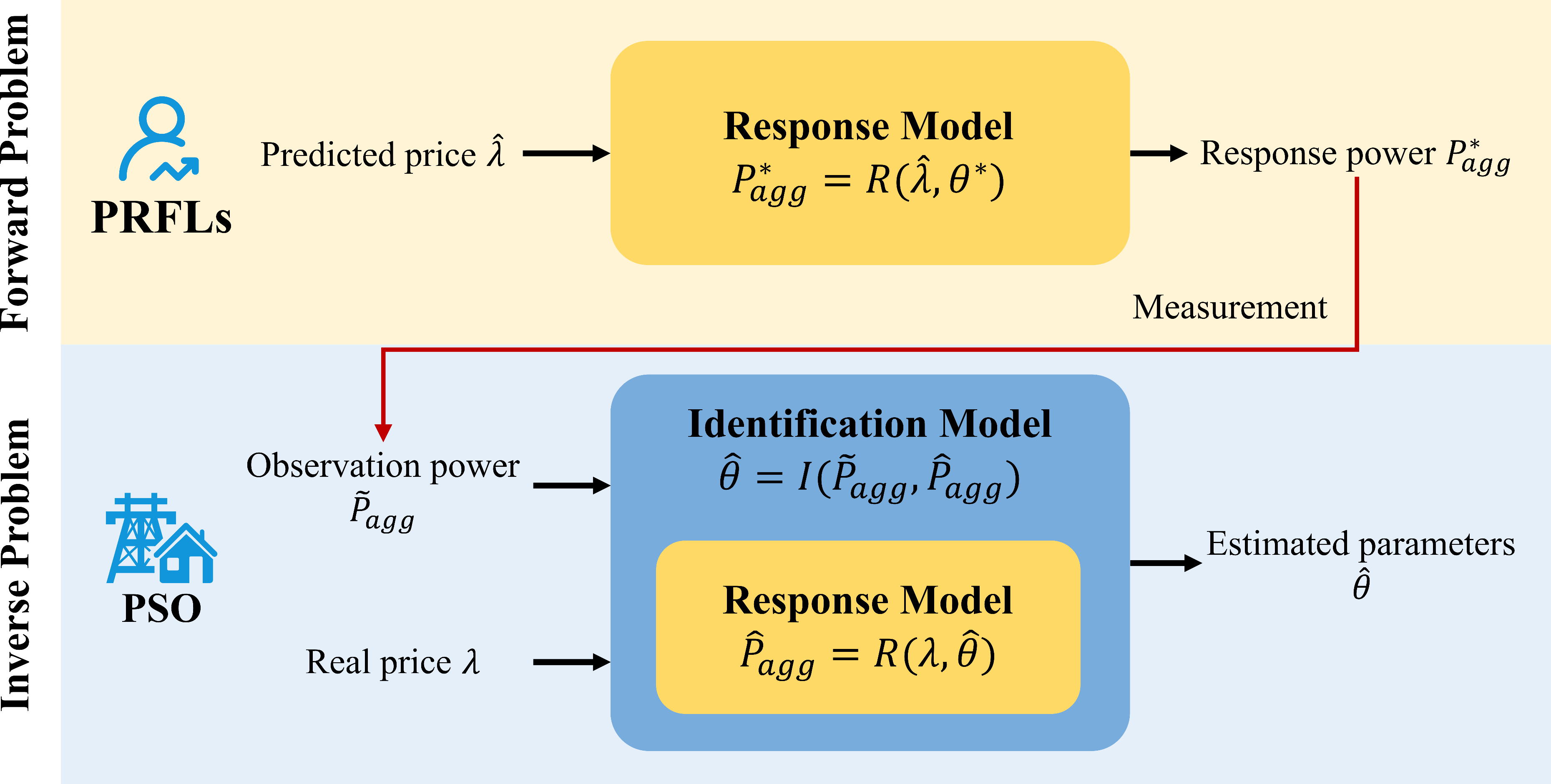}
\caption{Data-driven modeling framework of PRFLs.}
\label{IO}
\end{figure}

The identification process constitutes a typical IO framework as shown in Fig. \ref{IO}: 
(1) Forward problem: PRFLs solve the response model $P_{agg}^*=R(\hat\lambda,\theta^*)$ using forecast prices $\hat\lambda$ and actual AGM parameters $\theta^*$ within $\mathbf{\Omega}_T$ to optimize economic objectives. The actual aggregate response power $P_{agg}^*$ is observed by PSOs as $\tilde{P}_{agg}$ with measurement noise.
(2) Inverse problem: PSOs estimate parameters $\hat{\theta}$ in the surrogate AGM  $\hat{\mathbf{\Omega}}_T$ via identification model $\hat\theta=I(\tilde{P}_{agg},\hat{P}_{agg})$ under actual prices $\lambda$. The surrogate AGM mirrors the forward structure with distinct parameters, receiving true price signals rather than forecasts. Given that the response model is typically formulated as an optimization problem, this inverse problem inherently constitutes a bilevel program. Subsequent sections detail both problem formulations.

\begin{remark}
The time-invariant nature of  $\theta$ (constant power/energy bounds, storage efficiency, and ramp rates) inherently restricts the AGM applicability to systems with stable operational parameters during planning horizons. 
\end{remark}

\section{Forward Problem: Response Model }
\label{section3}
This section presents the PRFLs response model, considering both static and dynamic decision mechanisms. The planning horizon $T_{ph}$ and decision cycle $T_{dc}$ govern energy scheduling, wherein static mechanism employs $T_{dc}$-horizon price forecasts and $T_{dc}=T_{ph}$, while dynamic mechanism implements rolling horizon price updates and $T_{dc}<T_{ph}$. The core distinction lies in real-time price forecast integration. First, we introduce two fundamental assumptions in formulations, as:

\newtheorem{assumption}{Assumption}
\begin{assumption}
\label{as1}
The PRFLs are rational and optimize their response power to minimize their total economic costs.
\end{assumption}
\begin{assumption}
\label{as2}
The response of the PRFLs does not affect the electricity prices.
\end{assumption}

\vspace{-0.5cm}
\subsection{Static Response Model}
We assume that under the predicted price $\hat{\lambda} \in \mathbb{R}^{T_{dc}}$, the PRFLs utilize the following response model to determine the response power:
\begin{subequations}
\label{static_response_model}
\begin{equation}
\begin{aligned}
R_{st}:\quad  \hat{P}_{agg}(\theta,\hat{\lambda})=\mathop{\arg\min} \limits_{P_{agg} \in \mathbb{R}^{T_{dc}}} \; h_{st}(\theta) = \hat{\lambda}^{\top} P_{agg}
\end{aligned}
\end{equation}
\begin{equation}
s.t. \; P_{agg} \in \Omega_{T_{dc}}(\theta) \cap \{P_{agg} \left |[E_{str}^{i}]_{T_{dc}}=[E_{str}^{i}]_{0}\right. \},
\end{equation}
\end{subequations}
The supplementary constraint enforces equal initial and final energy storage levels within an operational cycle. Here, we use the predicted prices $\hat{\lambda}$ of the PRFLs instead of $\lambda$ in the response model  due to PRFLs' limited price visibility pre-decision. In the TOU pricing, $\hat{\lambda}=\lambda$ holds as prices are predetermined for $T_{ph}$. In real-time pricing implementations wherein prices are post-dispatched, $\hat{\lambda}= \lambda + \Delta\lambda$ with $\Delta\lambda$ representing forecast errors. 

\vspace{-0.3cm}
\subsection{Dynamic Response Model}
In the dynamic model, the PRFLs determine the response power for the sequential $T_{ph}$ periods and only implement the first $T_{dc}$ periods decisions. In this setting, the response model is formulated as:
\begin{subequations}
\label{dynamic_response_model}
\begin{equation}
\begin{aligned}
R_{dy}:\quad&\hat{P}_{agg}^{}(\theta, \hat{\lambda}_{dy}) =\mathop{\arg\min} \limits_{{P}_{agg} \in \mathbb{R}^{T_{dc}}} \; h_{dy}(\theta) \\
&= \hat{\lambda}_{dy}^\top P_{agg} -[\hat{\lambda}_{ave}]_{t} \sum_{i \in \mathbf{I}_{str}}e_{str}^{i,t+T_{dc}},\forall t\in \mathbf{T}, \\
\end{aligned}
\end{equation}
\begin{equation}
s.t. \; P_{agg} \in \mathbf{\Omega}_{T_{dc}}(\theta),   
\end{equation}
wherein 
\begin{equation}
\begin{aligned}
    &\hat{\lambda}_{dy}=[[ \hat{\lambda}_{dy}]_t,[\hat{\lambda}_{dy}]_{t+1},\dots,[\hat{\lambda}_{dy}]_{t+T_{dc}-1}]^\top,\\
   &[\hat{\lambda}_{ave}]_{t}={{\sum\limits_{{i} = {T_{dc}} + t}^{{T_{ph}} + t - 1} [\hat{\lambda}_{dy}]_{i}  }} / ({{{T_{ph}} - {T_{dc}}}}),\forall t\in \mathbf{T},
\end{aligned}
\end{equation}
\end{subequations}
wherein the last term of the objective is employed to assess the prospective value of energy storage. Logically, a higher potential value  will incentivize aggregators to augment the volume of energy stored. Conversely, a lower potential value will lead them to prioritize utilizing stored energy to meet demand rather than procuring electricity from the grid.

\vspace{-0.3cm}
\subsection{Extending to Integrating Inner Costs}
In some cases, besides the electricity-exchange costs defined in the models \eqref{static_response_model} and \eqref{dynamic_response_model}, the PRFLs also have the inner costs, such as the fuel cost and the loss of energy satisfaction. Recall the model $\mathbf{\Omega}_T$ defined in \eqref{eq_sum_model_T}, the inner costs are usually related to the  adjustable power $P_{adj}^i$. For example, if $P_{adj}^i$ denotes the power of one distributed generator or the interruptible load, then the fuel cost or the loss of energy satisfaction can be modeled as a quadratic function derived from the offset price and the power deviation to the expected power $\check{P}_{adj}^j$. Based on this, the functions $h_{st}(\theta)$ and $h_{dy}(\theta)$ in the models \eqref{static_response_model} and \eqref{dynamic_response_model} can be separately updated as:

\begin{subequations}
\begin{equation}
    h_{st}(\theta)=\hat{\lambda}^{\top} P_{agg}+\sum_{j \in \mathbf{I}_{adj}} c^j\Vert P_{adj}^j-\check{P}_{adj}^j \Vert_2^2,\\ 
\end{equation}
\begin{equation}
\begin{aligned}
       h_{dy}(\theta) &=\hat{\lambda}_{dy}^\top P_{agg} -[\hat{\lambda}_{ave}]_{t} \sum_{i \in \mathbf{I}_{str}}e_{str}^{i,t+T_{dc}} \\ 
    &+ \sum_{j \in \mathbf{I}_{adj}} c^j\Vert P_{adj}^j-\check{P}_{adj}^j \Vert_2^2,
\end{aligned}
\end{equation}
\end{subequations}
wherein the term related to price $\hat{\lambda}$ denotes the power purchase cost and the term containing $c^i$ means the inner cost. The $\check{P}_{adj}^j$ represents the expected power of the users.

\vspace{-0.2cm}
\section{Inverse Problem: Identification Model}
\label{section4}
To identify the parameters in the AGM of the PRFLs, we first formulate the identification model and then analyze the impact of noise on the identification results.

\vspace{-0.2cm}
\subsection{Identification Model}
Accurate assessment of PRFLs behavioral identification necessitates an objective function quantifying model deviations. Identification precision directly correlates with alignment between estimated $\hat{P}_{agg}$ and observed $\tilde{P}_{agg}$ responses under identical prices $\hat{\lambda}$. We thus formulate the accuracy metric using the $l_2$-norm:
\begin{equation}
\label{IO_static}
\begin{gathered}
I_{st}:\  {\hat \theta} = \mathop {\arg \min }\limits_\theta  f(\theta ) =  \sum_{n \in \mathbf{N}}\frac{1}{|\mathbf{N}|} \| {\tilde{P}_{agg}^n - \hat{P}_{agg}^n(\theta )} \|_2^2\\
s.t.\; \hat{P}_{agg}^n(\theta ) = \mathop {\arg \min }  \limits_{{P_{agg}^n} \in \Omega_{T_{dc}}(\theta) \cap \{[E_{str}^{i}]_{T_{dc}}=[E_{str}^{i}]_{0}  \}} \ h_{st}(\theta) ,\\
\forall n\in \mathbf{N}.
\end{gathered}\; 
\end{equation}
wherein \( \mathbf{N} \) represents the index set of the samples; $\tilde{P}_{agg}^n$ denotes the n-th measurement of the response power.

In a similar manner, this concept can be utilized to develop a model for the dynamic case, as:
\begin{equation}
\label{IO_dy}
\begin{gathered}
I_{dy}:\     {\hat{\theta}} = \mathop {\arg \min }\limits_\theta  f(\theta ) =  \sum_{n \in \mathbf{N}}\frac{1}{|\mathbf{N}|} \| {\tilde{P}_{agg}^n - \hat{P}_{agg}^n(\theta )} \|_2^2,\\ 
    s.t.\; [\hat{P}_{agg}^{n}(\theta )]_{t} = \mathop {\arg \min }\limits_{{P_{agg}^n}  \in \Omega_{T_{dc}}(\theta)} \ h_{dy}(\theta), \\
\forall t \in \mathbf{T}, n\in \mathbf{N}.
\end{gathered}
\end{equation}
The static model enforces periodic decision-making with energy recovery constraints, whereas the dynamic model implements rolling horizon optimization. This fundamental operational distinction yields divergent constraint formulas.

\vspace{-0.5cm}
\subsection{Impact of Noise}
In this following, we first model the noise in observation response power $\tilde{P}_{agg}$ and predicted fixed component power $\hat{P}_{fix}$. Based on that, we further propose Theorem \ref{theory1} on the noise impact on the solution to the inverse problem in static case.

To characterize physical uncertainties, we model the noise in fixed component prediction and power observation as $\varepsilon_{fix}\sim \mathcal{N}(\mu_{fix},\Sigma_{fix})$ and $\varepsilon_{agg} \sim \mathcal{N}(\mu_{agg},\Sigma_{agg})$ respectively, wherein $\mu_{(\cdot)}\in \mathbb{R}^T$ and $\Sigma_{(\cdot)}\in\mathbb{R}^{T\times T}$ denote mean vector and covariance matrix of Gaussian distribution. The noise data can be decompose as:
\begin{equation}
\label{p_fixed}
\hat{P}_{fix}=P_{fix}^*+\varepsilon_{fix},\ 
\tilde{P}_{agg}={P}_{agg}^*+\varepsilon_{agg}.
\end{equation}

Here $P_{fix}^*$ and ${P}_{agg}^*$ represent the actual power values, with $\varepsilon_{fix}=[[\varepsilon_{fix}]_1,[\varepsilon_{fix}]_2,\cdots,[\varepsilon_{fix}]_T]^\top$ and $\varepsilon_{agg}=[[\varepsilon_{agg}]_1,[\varepsilon_{agg}]_2,\cdots,[\varepsilon_{agg}]_T]^\top$ being prediction/observation noise vectors.
The identification objective function derives from noise data as:
\begin{equation}
    \begin{aligned}
f_{nd}(\theta)&=\frac{1}{\left |\mathbf{N} \right |}  \sum_{n \in \mathbf{N}}\left \| \tilde{P}_{agg}^n-\hat{P}_{agg}^n\right \|^2 \\
&=\frac{1}{\left |\mathbf{N} \right |}  \sum_{n \in \mathbf{N}}\left \| {\Delta}^n+\varepsilon^n_{agg}- \varepsilon_{fix}^n\right \|^2  \\
&=\frac{1}{\left |\mathbf{N} \right |}  \sum_{n \in \mathbf{N}}\;\left \| {\Delta}^n+\mu_{agg}- \mu_{fix} \right \|^2\\
&+\frac{2}{\left |\mathbf{N} \right |}  \sum_{n \in \mathbf{N}} (\overline{\Delta}^n)^\top(\varepsilon^n_{agg}- \varepsilon_{fix}^n-\mu_{agg}+\mu_{fix})\\
&+\frac{1}{\left |\mathbf{N} \right |}  \sum_{n \in \mathbf{N}}(\left \|\varepsilon^n_{agg}- \varepsilon_{fix}^n\right \|^2 -\left \| \mu_{agg}- \mu_{fix}\right \|^2).
\label{obj_noise}
    \end{aligned}
\end{equation}

It is worth noting that the objective function for noiseless data is a special case of formula \eqref{obj_noise}, as:
\begin{align}
&f_{nf}(\theta) =\frac{1}{\left |\mathbf{N} \right |}  \sum_{n \in \mathbf{N}}\;\left \| {\Delta}^n \right \|^2,
\label{obj_nf}
\end{align}
wherein $\Delta^n=P_{str}^{n*}-\hat{P}_{str}^n+P_{adj}^{n*}-\hat{P}_{adj}^n$ and $\overline{\Delta}^n={\Delta}^n+\mu_{agg}- \mu_{fix}$ represent the deviation in deterministic part.

Based on the characteristics of the above objective function, we propose the Theorem \ref{theory1} on the noise effect of the solution.
\begin{theorem}
\label{theory1}
    If the $l^2$-norm is used for $f_{nd}(\theta)$ in model \eqref{obj_noise} and $\forall n\in \mathbf{N}, \varepsilon^n_{agg}- \varepsilon_{fix}^n\sim\mathcal{N}\left(0,\Sigma_P\right) (\Sigma_P\in\mathbb{R}^{T\times T}$ is the covariance matrix), we have:
\begin{flalign*}
&\ (a)\lim_{\left|\mathbf{N}\right|\rightarrow+\infty}\left({\hat\theta}_{nd}-{{\hat\theta}}_{nf}\right)=0, \\
&\ (b)(f_{nd}({\hat\theta})-f_{nf}({\hat\theta}))_N{\stackrel{Probility}{\longrightarrow}}\mathrm{tr}\left(\Sigma_P\right),&
\end{flalign*}
wherein the ${\hat{\theta}}_{nd}$ and ${\hat{\theta}}_{nf}$ represent the optima for models \eqref{obj_noise} and \eqref{obj_nf} respectively, and $\mathrm{tr}(\cdot)$ denote the trace of matrix, with $\mathrm{tr}\left(\Sigma_P\right)=\mathrm{tr}\left(\Sigma_{agg}+\Sigma_{fix}\right)$. A detailed proof is given in the Appendix \ref{proof1}.
\end{theorem}

\begin{remark}
    The mean of observation and prediction noise will results in a deviation from the optimal response power, i.e., the actual response power $P_{agg}^{n*}$ will be observed as $P_{agg}^{n*}+\mu_{agg}-\mu_{fix}$. This will result in differing levels of drift in our estimated parameters $\hat{\theta}$.  Fortunately, in practical engineering, the mean of the noise can be regarded as 0, i.e., $\mu_{agg}=0,\mu_{fix}=0$. Therefore, the estimated parameters $\hat{\theta}$ will be mildly affected when the train set is large enough.
\end{remark}
\begin{remark}
     While the constant term $\mathrm{tr}(\Sigma_{agg})+\mathrm{tr}(\Sigma_{fix})$ does not influence the optimal solution for $\hat{\theta}$, it does impact our assessment of the optimal solution. The value of $f_{nd}(\hat{\theta})$ will be always greater than the value in noise-free case even if the parameter $\theta$ is accurately estimated. Still, a smaller $f_{nd}(\hat{\theta})$ means a better estimation of the parameters $\theta$.
\end{remark}

\section{Bayesian Identification Algorithm}
\label{section5}
\begin{figure}[tb]
\centering
\footnotesize
\includegraphics[width=1\linewidth]{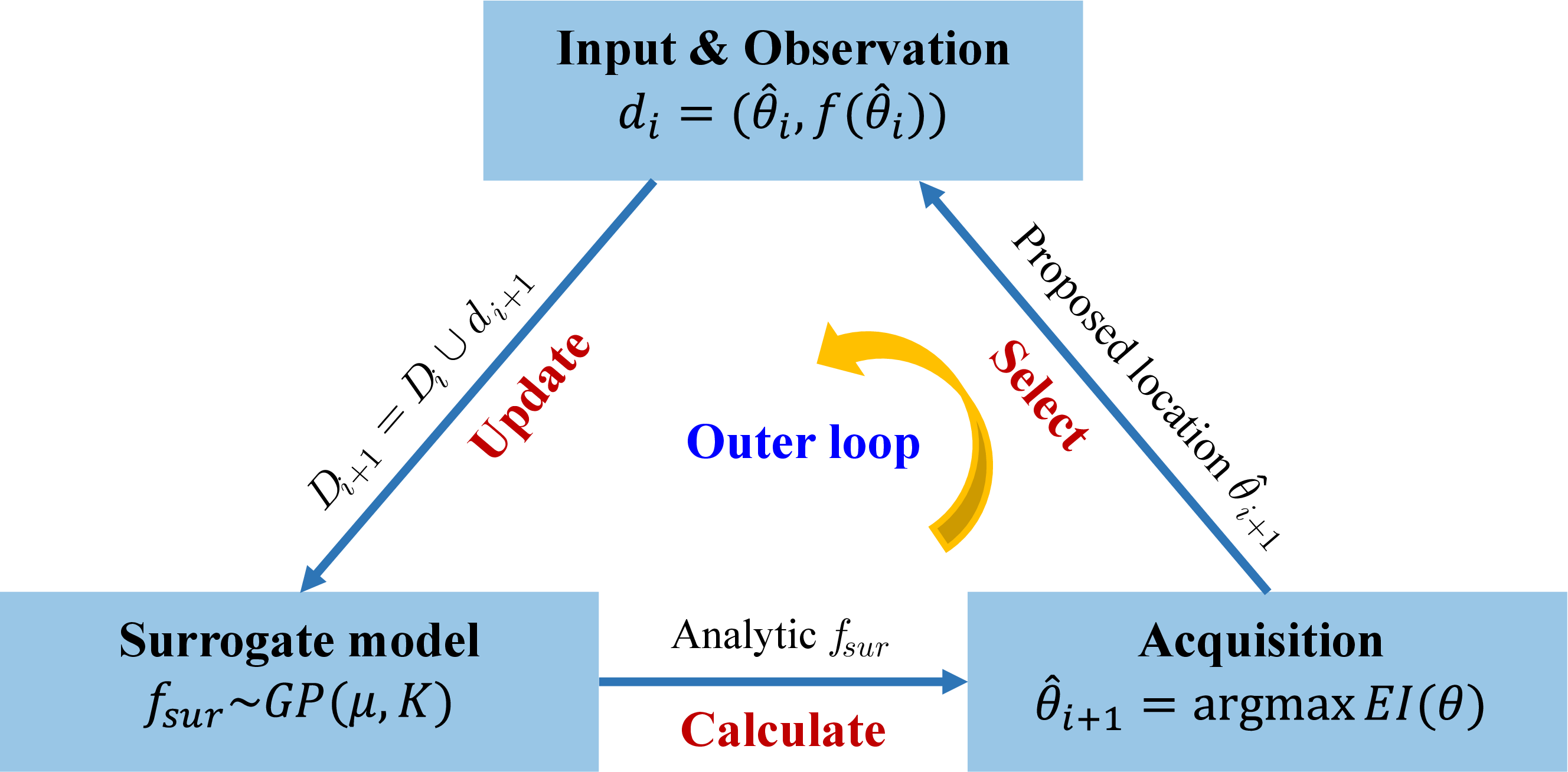}
\caption{Algorithm framework based on Bayesian loop.}
\label{Bayesian_loop}
\end{figure}

In this section, we first present the BayOpt-based solution framework for solving identification model, followed by a proposed recursive block Cholesky decomposition (BCD) method to accelerate covariance matrix inversion in the computational process. The developed algorithm features a bilevel structure comprising three core modules, as illustrated in Fig. \ref{Bayesian_loop}, which will be elaborated in the following subsections.

\vspace{-0.5cm}
\subsection{Solution Method}
The algorithm's core implements a Bayesian loop that iterates through three modules sequentially until reaching predefined cycle limits $N_{max}$.
\subsubsection{Observation module}
This module replicates the forward problem structure: given parameters $\theta_i$ at iteration $i$, it solves the response model $R(\hat\lambda,\theta_i)$ and computes the value of identification objective $f(\theta_i)$. The response model, formulated as a linear program, can be efficiently solved by commercial optimization solvers. Resultant pairs $d_i=(\theta_i,f(\theta_i))$ are then transferred to the surrogate model as training data.
\subsubsection{Surrogate model}
The BayOpt framework fundamentally differs from conventional optimization by constructing a probabilistic surrogate model through Gaussian process (GP) regression. Formally, the surrogate model $f_{sur}(\theta)\sim GP(\mu(\theta),k(\theta_i,\theta_j))$. In this work, based on the continuity of the objective function, we choose the Matern kernel \cite{Moriconi2020Highdimensional} as the covariance function, as:
\begin{equation}
\begin{aligned}
\label{kernal_function}
&k(\theta_i,\theta_j)=\mathbb{E}[f(\theta_i)f(\theta_j)]\\
&=\alpha (1+\frac{\sqrt{5}r}{\beta}+\frac{5r^2}{3\beta^2}){e}^{-\frac{\sqrt{5}r}{\beta}}+\epsilon^2\delta_{ij},
\end{aligned}
\end{equation}
wherein the hyperparameters $\eta=(\alpha,\beta,\epsilon)$ govern signal variance, length scale, and noise respectively. In each loop with small sample, we re-estimation GP hyperparameter through marginal likelihood maximization. The Kronecker delta $\delta_{ij}$ in the kernel function models heteroscedastic noise, critical for handling the uncertainties. 

For the samples $f_{sur}(\theta_{1:n}) := [f(\theta_1),f(\theta_2),\cdots ,f(\theta_n)]^\top$, the GP prior distribution becomes:
\begin{subequations}
\label{prior_GP}
\begin{equation}
f_{sur}(\theta_{1:n})\sim\mathcal{N}(\mu(\theta_{1:n}),K(\theta_{1:n},\theta_{1:n})),
\end{equation}
wherein
\begin{equation}
    \mu(\theta_{1:n})=[\mu(\theta_1),\mu(\theta_2),\cdots,\mu(\theta_n)]^\top, 
\end{equation}
\begin{equation}
K(\theta_{1:n},\theta_{1:n})=\begin{bmatrix}
k(\theta_1,\theta_1)&\cdots&k(\theta_1,\theta_n) \\ 
 \vdots & \ddots  & \vdots\\ 
k(\theta_n,\theta_1) & \cdots & k(\theta_n,\theta_n)
\end{bmatrix}.
\end{equation}
\end{subequations}

The covariance matrix $K\in \mathbb{R}^{n \times n}$ with elements $[K]_{ij}=k(\theta_i,\theta_j)$ captures the parameter space correlations. This nonparametric formulation enables sequential updating of posterior distributions $\mathbb{P}(f_{sur}(\theta)|f_{sur}(\theta_{1:n}))$ through Bayesian inference as:
\begin{align}
\label{posterior_GP}
&f_{sur}(\theta)|f_{sur}(\theta_{1:n})\sim\mathcal{N}(\mu_*(\theta),K_*) \notag\\
&\mu_*(\theta)=k(\theta_{1:n},\theta)^T K(\theta_{1:n},\theta_{1:n})^{-1}f(\theta_{1:n}) \notag\\
&K_*(\theta)=k(\theta,\theta)- k(\theta_{1:n},\theta)^TK(\theta_{1:n},\theta_{1:n})^{-1}k(\theta_{1:n},\theta) \notag
\end{align}

\subsubsection{Selection module}
The GP is updated iteratively by incorporating additional observation points. Therefore, the selection of new sample has a significant impact on the accuracy of posterior GP. The selection module deduce the best location in parameter space via the acquisition function which serves as the valuation criterion of the next sampling point relying on the current posterior GP. Specifically, to propose the best new sample, this module should balance exploration-exploitation tradeoffs. In this paper we choose the expected improvement (EI) as the acquisition function to reduce the calculation of this module, expressed in formula \eqref{EI} 
\begin{equation}
\begin{aligned}  
\label{EI}
    EI(\theta)&:=\mathbb{E}([f_{sur}^*-\hat{f}_{sur}(\theta)]^+),
    \end{aligned}
\end{equation}
wherein  $f_{sur}^*=\min\{f_{sur}(\theta_1),f_{sur}(\theta_2),\cdots,f_{sur}(\theta_n)\}$ represents the incumbent optimum. 
The closed-form solution under GP prior assumptions enables efficient global optimization for EI, as:
\begin{equation}
\theta_{n+1}=\arg\max_{\theta }{((f_{sur}^*-\mu_*(\theta))\Phi(z)+K_*(\theta)\phi(z))}, 
\label{EI_solve}
\end{equation}
wherein $z=\frac{f^*-\mu(\theta)}{K_*(\theta)}$, with $\Phi(\cdot)$ and $\phi(\cdot)$ denoting the standard normal CDF and PDF. 
It is noted that the observation module updates the dataset iteratively by incorporating new samples $d_{n+1}$ from the selection module, such that $D_{n+1}=D_n\cup d_{n+1}=(\theta_{1:n},f(\theta_{1:n}))\cup (\theta_{n+1},f(\theta_{n+1}))$. While expanding the dataset enhances surrogate model accuracy, it simultaneously increases computational demands. Thus, the cycle limit number $N_{max}$ represents a critical compromise between estimation precision and computational feasibility.


\vspace{-0.4cm}
\subsection{Acceleration Method}
\label{FABO}
\begin{algorithm}[tb]
\footnotesize
\label{FABO_algorithm}
    \SetAlgoLined 
	\caption{Parameter estimation based on FABO}
	\KwIn{$\theta_{1:n_0},f,N_0,N_{max}$}
	Calculate the objective function value of $\theta_{1:n_0}$ to obtain the initial data $D_{n_0}$ and set the condition number of classic loop $N_0$ and that of entire loop $N_{max}$\;
        Tune hyperparameters by solve $\eta^*= \arg\max_{\eta}{p(f_{sur}(\theta_{1:n_0})|\eta,\theta_{1:n_0})}$, then calculate $K(\theta_{1:n_0},\theta_{1:n_0}),L_{n_0}$ and set $n=n_0$\;
	\While{$n\leq N_0$}{
        Calculate $J_{1}  = {L_{n}^{ - 1}}f_{sur}({\theta _{1:n}}),J_2  = {L_{n}^{ - 1}}k({\theta _{1:n}},\theta )$\;
        Solve $\theta_{n+1}=\arg\max_{\theta }{EI(\theta)}$ with calculating $\mu_*,K_*$ by $J_{1},J_{2}$ in each iteration\;
        Obtain the new data $d_{n+1}=(\theta_{n+1},f(\theta_{n+1}))$ by solving forward problem \;
        Tune hyperparameters again on new data by solve $\eta^*= \arg\max_{\eta}{p(f_{sur}(\theta_{1:n+1})|\eta,\theta_{1:n+1})}$ \;
        $n=n+1$\;
        }
        Save the matrix $L_{N_0}^{-1}$ and last hyperparameters $\eta$ in above loop\;
        \While{$N_0+1\leq n\leq N_{max}$}{
        Calculate the inversion of $L_{n}$ by BCD\;
        Solve $\theta_{n+1}=\arg\max_{\theta }{EI(\theta)}$ with calculating $\mu_*,K_*$ using $L_n^{-1}$ in each iteration\;
        Obtain the new data $d_{n+1}=(\theta_{n+1},f(\theta_{n+1}))$ by solving forward problem\;
        $n=n+1$\;
        }
		Return the optimal point in result set $\theta^*=\argmin_{(\theta,f(\theta)) \in D_n}{f(\theta)}$.
        \label{Algoritm 1}
\end{algorithm}
It is well known that the cubic-time complexity $O(n^3)$ inherent in Bayesian optimization primarily stems from the inversion of covariance matrices in formula $K_*$ – a fundamental computational bottleneck governed by Cholesky decomposition requirements in kernel-based regression. 
To address the computational bottleneck in covariance matrix inversion, we develop the BCD method to update the matrix leveraging the information of low-rank matrix. Specifically, Theorem \ref{lemma} provides the calculation method with computational complexity $O(n^2)$ of this problem: 
\begin{theorem}
\label{lemma}
Denote $L_n$ as the Cholesky decomposition of covariance matrix $K_n$ at $n$ iteration, satisfying $K_n=L_nL_n^{\top}$. Given $L_n$, $L_n^{-1}$, and $\theta_{1:n+1}$, the updated decomposition $L_{n+1}$ and its inversion $L_{n+1}^{-1}$ for augmented covariance matrix $K_{n+1}$ are calculated as:
\begin{subequations}
\begin{equation}
K_{n + 1} = K_{n + 1,n + 1} - (L_{n + 1}^{ - 1}{K_{n + 1,1}})^TL_{n + 1}^{ - 1}K_{n + 1,1},
\end{equation}
\begin{equation}
{{L_{n + 1}} = \left[ {\begin{array}{*{20}{c}}
{{L_n}}&0\\
{{L_{21}}}&{{L_{22}}}
\end{array}} \right],L_{n + 1}^{ - 1} = \left[ {\begin{array}{*{20}{c}}
{L_1^{ - 1}}&0\\
{{L_3}^{ - 1}}&{L_2^{ - 1}}
\end{array}} \right]},
\end{equation}
\end{subequations}
\begin{align}
 &\mathrm{wherein} \ 
    {L_{21}}= K_{n + 1,1}^ \top L_n^{ -  \top },
    {L_{22}} = \sqrt {{K_{n + 1,n + 1}} - {L_{21}}L_{21}^ \top },\notag \\
&L_1^{ - 1} = L_n^{ - 1},
    L_2^{ - 1} = \frac{1}{L_{22}},
    L_3^{ - 1} =  - L_{22}^{ - 1}K_{n + 1,1}^ \top {({L_n}L_n^ \top )}^{ - 1}.\notag \\
&K_{n + 1,n+1}=k(\theta_{n+1},\theta_{n+1}),\notag \\
&K_{n + 1,1}=[k(\theta_1,\theta_{n+1}),k(\theta_2,\theta_{n+1}),\cdots,k(\theta_n,\theta_{n+1})]^{\top}.\notag 
\end{align}

\end{theorem}

The developed recursive BCD calculation method achieves $O(n^2)$ computational complexity in formula $L_3^{-1}$ per iteration. The detailed proof is given in Appendix \ref{proof2}. This complexity reduction stems from the analytical inverse propagation through partitioned matrix identities, effectively avoiding full covariance matrix recomputation while maintaining numerical precision.

\section{Numerical Tests}
\label{section6}
The numerical test considers an AGM comprising a 400-600 kW fixed load, 10 adjustable power components, consisted of 5 distributed generators and 5 interruptible loads, and 50 energy storage systems. These generators' operational constraints contain 3-5 kW lower power bounds, 10-15 kW upper power bounds and 2-3 kW/h ramp rates. The interruptible loads are limited by 1-3 kW lower power bounds and 5-10 kW upper power bounds. The energy storage systems' parameters including the power bounds, energy capacity and the storage efficiency, sampling  from following uniform distribution $U(\cdot,\cdot)$: (1) Power bounds:  $ \underline{p}_{str}^i\sim U(-18,-6)$kW, \ $\overline{p}_{str}^i\sim U(4,16)$kW; 
(2) Energy capacity: $\overline{e}_{str}^i\sim U(8,64)$kWh, $\underline{e}_{str}^i \sim U(0.1\overline{e}_{str}^i,0.15\overline{e}_{str}^i)$kW; 
(3) Power bounds:   $\underline{p}_{str}^i\sim U(-18,-6) kW$.
The temporal framework adopts a 24-hour planning horizon with 1-hour resolution. Price signals $\lambda$ derive from ISO-NE 2022 market day-ahead prices (NEMA region) for static response model and real-time prices for dynamic response model. 

In the follow, we first evaluate the accuracy of the proposed BayOpt based method for the noisy static model, and compare the performance under different numbers of energy storage components. Then, we test the identification of the extended static model in the same way and analyze the effect of noise on the results. In addition, the Gaussian regression model outputs the relation between the loss function and each parameter, which helps us to analyze the identifiability of each parameter. In the same way, we display the deviation of the estimated aggregate response power $\hat{P}_{agg}$ in dynamic response mechanisms. We further compare the performance with the mainstream Newton-based approach. All of the above simulation experiments are run on a computer with an i7-14700k CPU using MATLAB R2020b software.

\vspace{-0.3cm}
\subsection{Performance for Static Response Model}
\label{res_static}
\begin{figure}[tb]
\centering
\footnotesize
\includegraphics[width=1\linewidth]{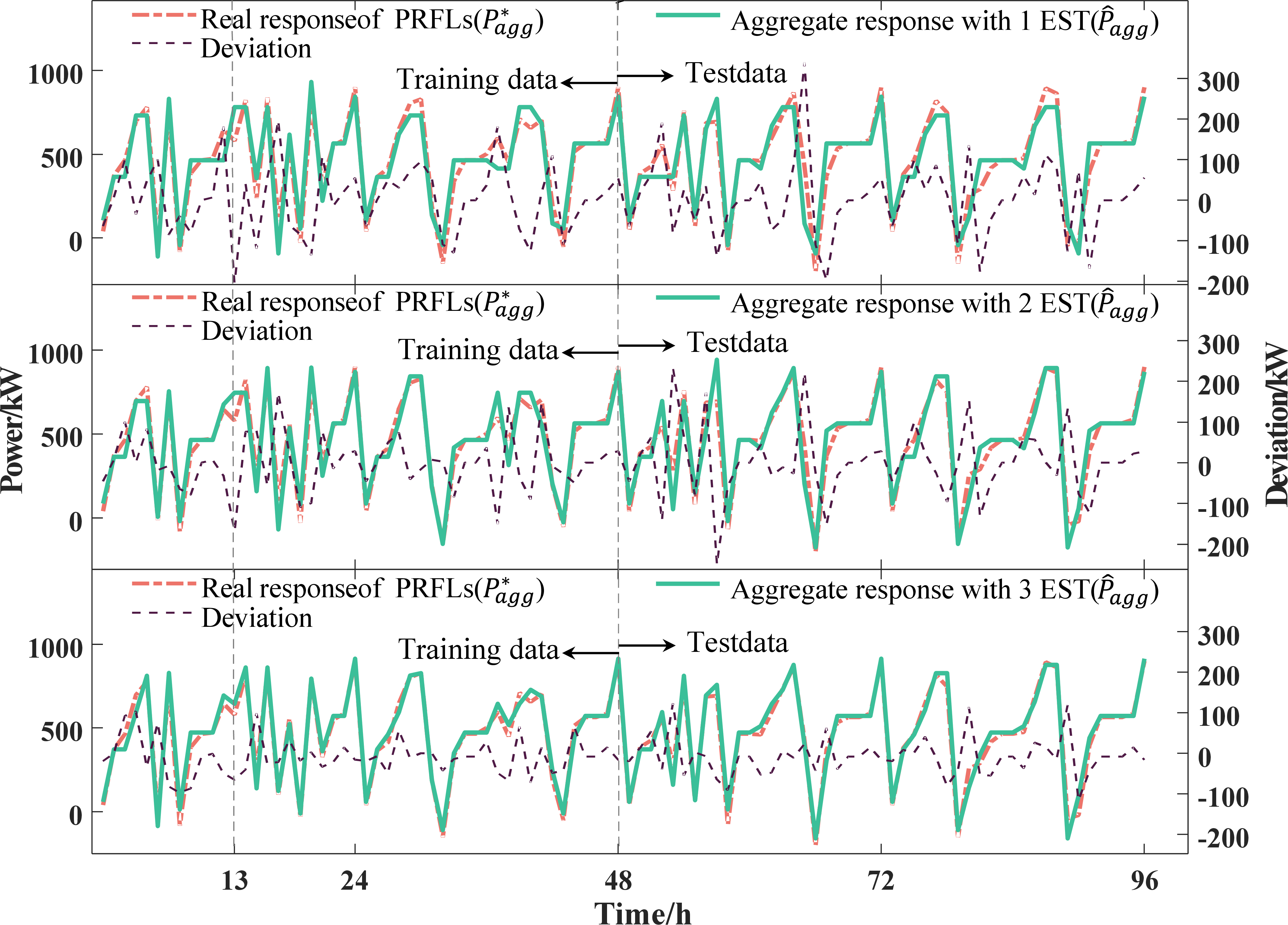}
\caption{Performance of  static model with noise data.}
\label{static_case}
\end{figure}

\begin{figure}[t]
\centering
\footnotesize
\includegraphics[width=1\linewidth]{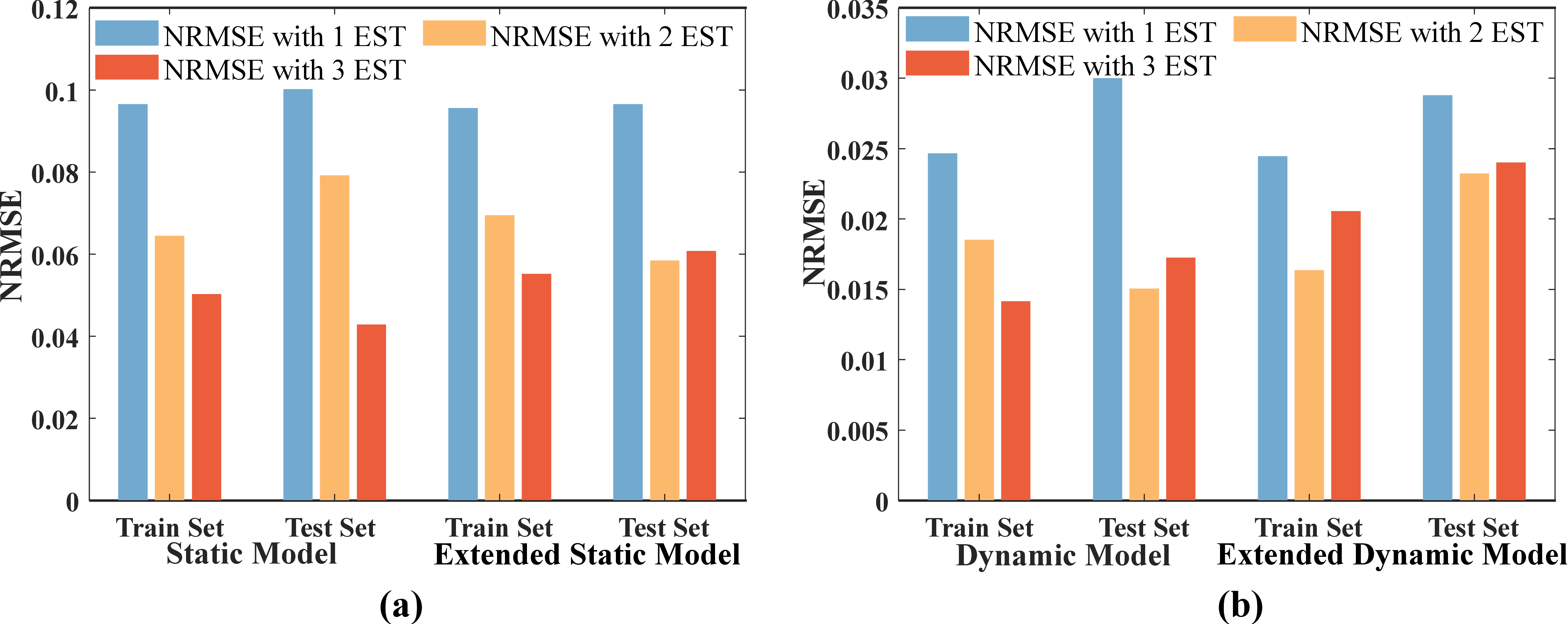}
\caption{NRMSE under different number of energy storages: (a) (Extended) Static case; (b) (Extended) Dynamic case.}
\label{NRMSE_static_dynamic}
\end{figure}

This experimental workflow comprises four stages:(1) Generate the ground-truth response \( P^*_{agg} \) by solving static response model \eqref{static_response_model} with the actual parameters in AGM; (2) Synthesize noisy training data \( \widetilde{P}_{agg} =P^*_{agg}+\epsilon\), wherein $\epsilon_t\sim\mathcal{N}(0,0.005[P_{agg}^{*}]_t)$, to simulate the noise introduced in measurement and prediction process; (3) Identify parameters $\hat\theta_{nf}$ and $\hat\theta_{nd}$ in AGM with different number of energy storages within surrogate AGM using pristine/noisy training pairs $(\lambda^k,P_{agg}^{k*})/(\lambda^k, \tilde{P}_{agg}^k)$ via BayOpt based algorithm;(4) After solving the response model with identified parameters, we evaluate test-set performance via aggregate response deviation $||P^*_{agg}-\hat{P}_{agg}||_2$ as shown in the Fig. \ref{static_case} and compare the error index illustrated in Fig. \ref{NRMSE_static_dynamic}. The error index NRMSE are calculated as:
\begin{equation}
NRSME=\frac{{\sqrt {\frac{1}{{T|\mathbf{N}|}}\sum_{n\in\mathbf{N}}{{\left\| {{P^{n*}_{agg}-\hat{P}_{agg}^n}} \right\|}_2^2}} }}{{\max_{n,t} \{ [P^{n*}_{agg}]_{t}\} }-{\min_{n,t} \{ [{P^{n*}_{agg}}]_{t}\} }}. 
\end{equation}

\begin{figure}[tb]
\centering
\footnotesize
\includegraphics[width=1\linewidth]{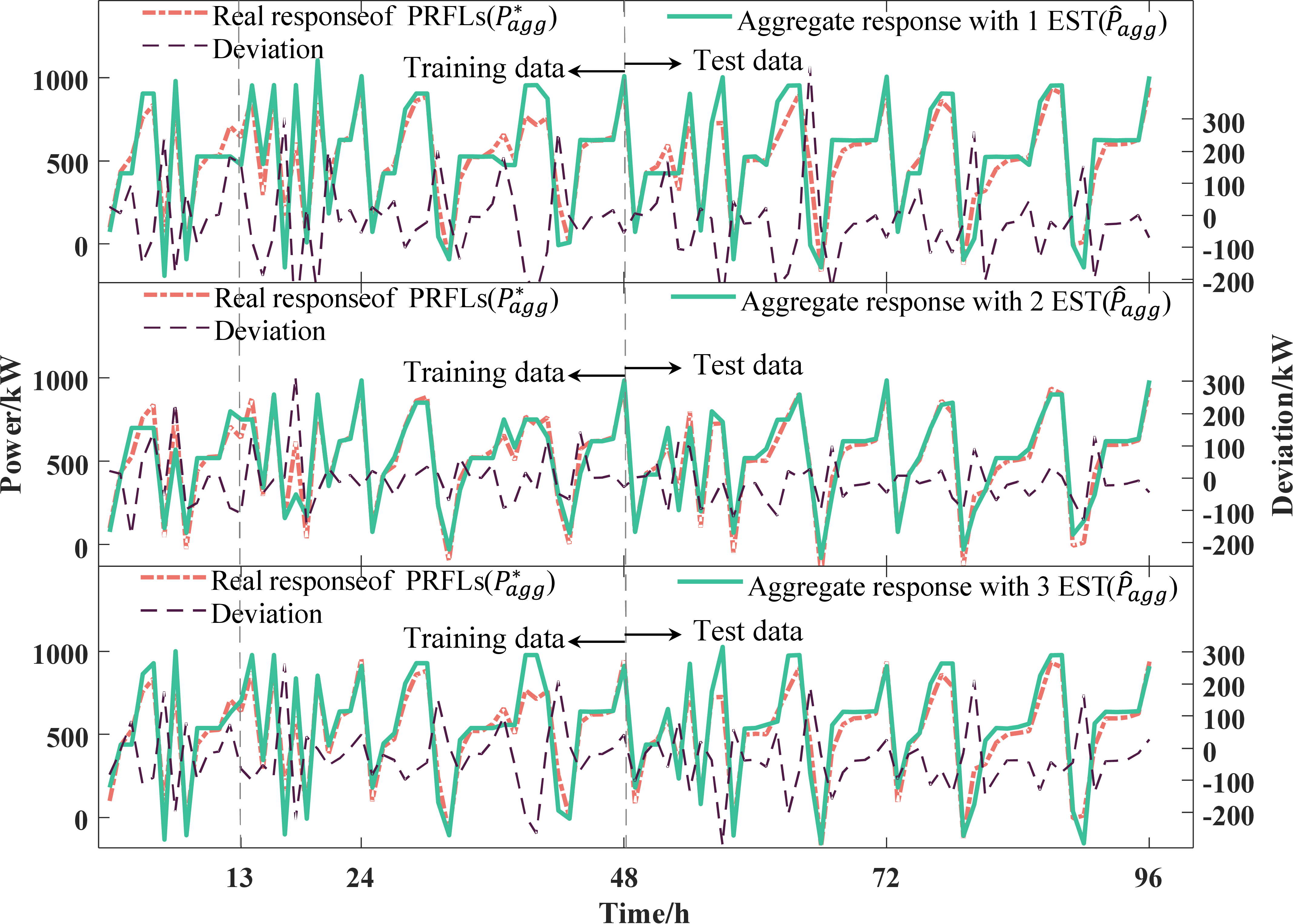}
\caption{Performance of  extended static model with noise data.}
\label{static_ex_case}
\end{figure}

It is worth noting that in Fig. \ref{static_case} and Fig. \ref{static_ex_case} at $t=13h$, as the number of energy storage increases, the response becomes more accurate, illustrating that the flexibility of the surrogate AGM can be enhanced by adjusting this number to match the flexibility of the real PRFLs, which can also be concluded in Fig. \ref{NRMSE_static_dynamic}(a). The  deviation in train set of extended model is bigger than that of the static model probably due to the complex quadratic response mechanism. 

\begin{figure}[tb]
\centering
\footnotesize
\includegraphics[width=1\linewidth]{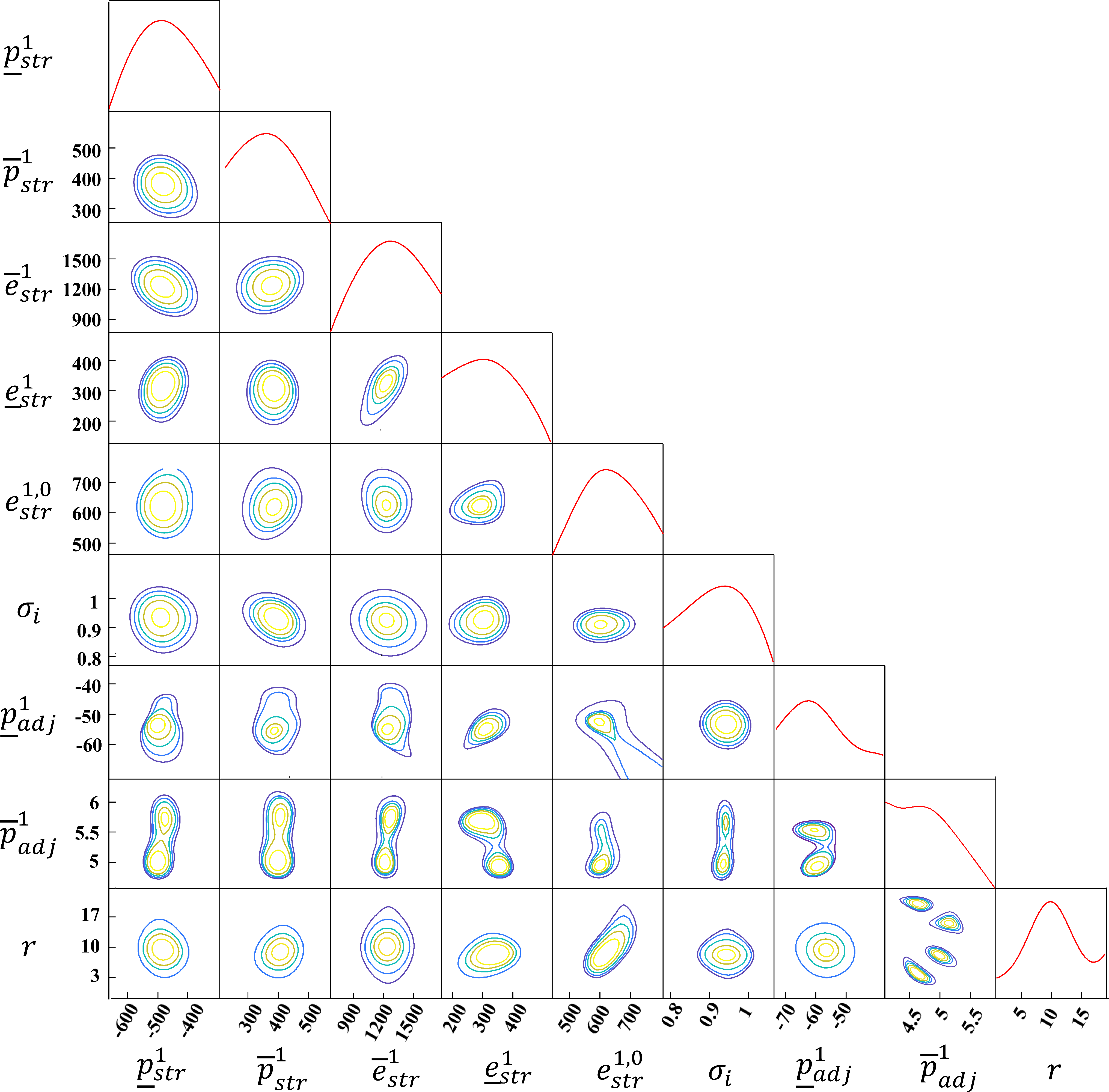}
\caption{Contour plots of parameters.}
\label{correlation}
\end{figure}

The Gaussian process regression analysis reveals  parameter identifiability, as visualized in Fig. \ref{correlation}.
For energy storage parameters, the parameter planes exhibit a singular extremum, demonstrating strong posterior identifiability. This unimodal distribution indicates robust convergence to unique solutions during parameter identification process.
Conversely, for adjustable power components, the rate-limiting parameters of adjustable power units display multiple extrema and limited parameter sensitivity ranges in the second-to-last line of pictures. These characteristics correspond to weak identifiability, particularly under static pricing conditions wherein positive electricity prices render such constraints always inactive. In the Fig. \ref{correlation}, we can further find an interesting results.
While the energy storage efficiency $\sigma^i$ emerges as the most influential parameter, current identification approaches are unable to identify this particular type of parameters. 

\begin{table}
\centering
\footnotesize
\belowrulesep = 0pt
\aboverulesep = 0pt
\caption{Noise effect on identified parameters.}
\begin{tabular*}{0.975\linewidth}{c|cc|cc|cc}
\toprule
\multirow{2}{*}{{$\hat\theta$}}& \multicolumn{2}{c|}{3 days} & \multicolumn{2}{c|}{10 days} & \multicolumn{2}{c}{55 days} \\ \cmidrule{2-7}
& $\theta_{nf}$ & $\theta_{nd}$ & $\theta_{nf}$ & $\theta_{nd}$& $\theta_{nf}$ & $\theta_{nd}$\\ 
\hline
$\underline{p}_{str}$ & -425.53 & -300.00 & 437.24 & 500.00 & 1343.10 & 1500.00\\
$\overline{p}_{str}$ & -566.86 & -582.56 & 366.57 & 358.72 & 1166.42 & 1146.80\\
$\overline{e}_{str}$ & -483.78 & -528.46 & 408.11 & 385.77 & 1270.27 & 1214.41\\
$\underline{e}_{str}$ & -437.10 & -610.43 & 431.45 & 344.79 & 1328.64 & 1111.96\\
${e}_{0}$ & -300 & -416.88 & 500& 441.56 & 1500.00 & 1353.90 \\
$\sigma$ & -457.57 & -453.36 & 421.21 &423.32 & 1303.03 & 1308.30\\
\hline
$\beta$ & \multicolumn{2}{c|}{62.11} & \multicolumn{2}{c|}{51.84} & \multicolumn{2}{c}{18.61}\\
\bottomrule
\end{tabular*}
\label{tab_noise_effect}
\end{table}

To verify the Theorem \ref{theory1}, we collect identified parameters when the number of samples increases from 3 days to 55 days under the static case with 1 EST in the Table \ref{tab_noise_effect}, wherein parameter units remain consistent with prior configurations. The parameter $\beta$ quantifies the discrepancy between parameter  derived from noise-corrupted and noise-free scenarios, expressed mathematically as: 
\begin{align}
\beta=||\theta^s_{nf}-\theta^s_{nd}||_2,
\end{align}
wherein
\(
\ \theta^s_{nf}=\frac{\theta_{nf}-\underline{\theta}_{nf}}{\overline{\theta}_{nf}-\underline{\theta}_{nf}},\ 
\theta^s_{nd}=\frac{\theta_{nd}-\underline{\theta}_{nd}}{\overline{\theta}_{nd}-\underline{\theta}_{nd}}
\).

Here $\underline{\theta}_{(\cdot)}$/$\overline{\theta}_{(\cdot)}$ represents the lower bounds and the upper bounds for identified parameters.
From the table \ref{tab_noise_effect}, we can evidently find the identification deviation $\beta$ decreases heavily with increasing sample size, consistent with the Theorem \ref{theory1}.

\vspace{-0.3cm}
\subsection{ Performance for Dynamic Response Model}
\label{res_dynamic}
\begin{figure}[tb]
\centering
\footnotesize
\includegraphics[width=1\linewidth]{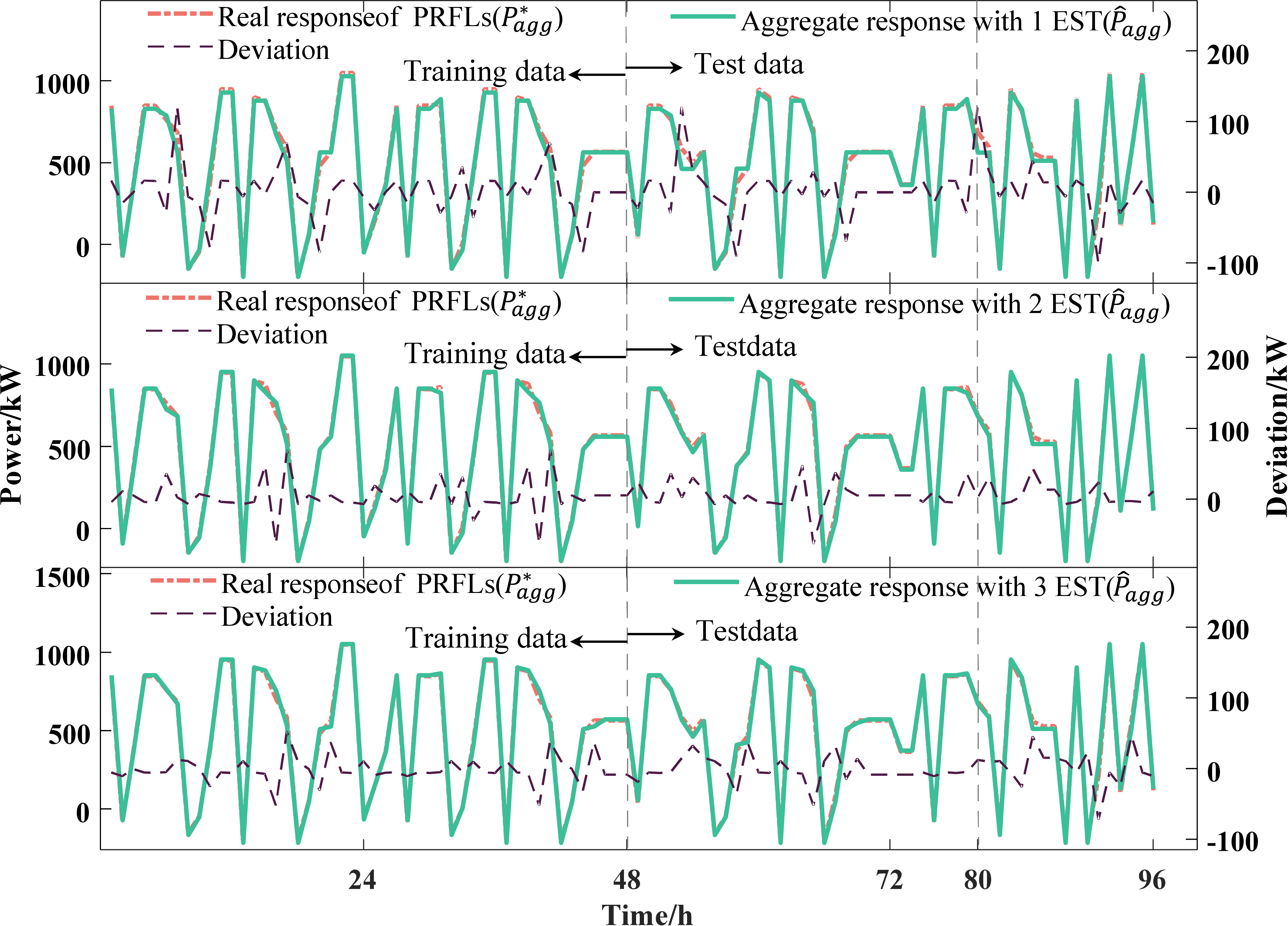}
\caption{Performance of  dynamic model with noise.}
\label{dynamic_case}
\end{figure}

\begin{figure}[tb]
\centering
\footnotesize
\includegraphics[width=1\linewidth]{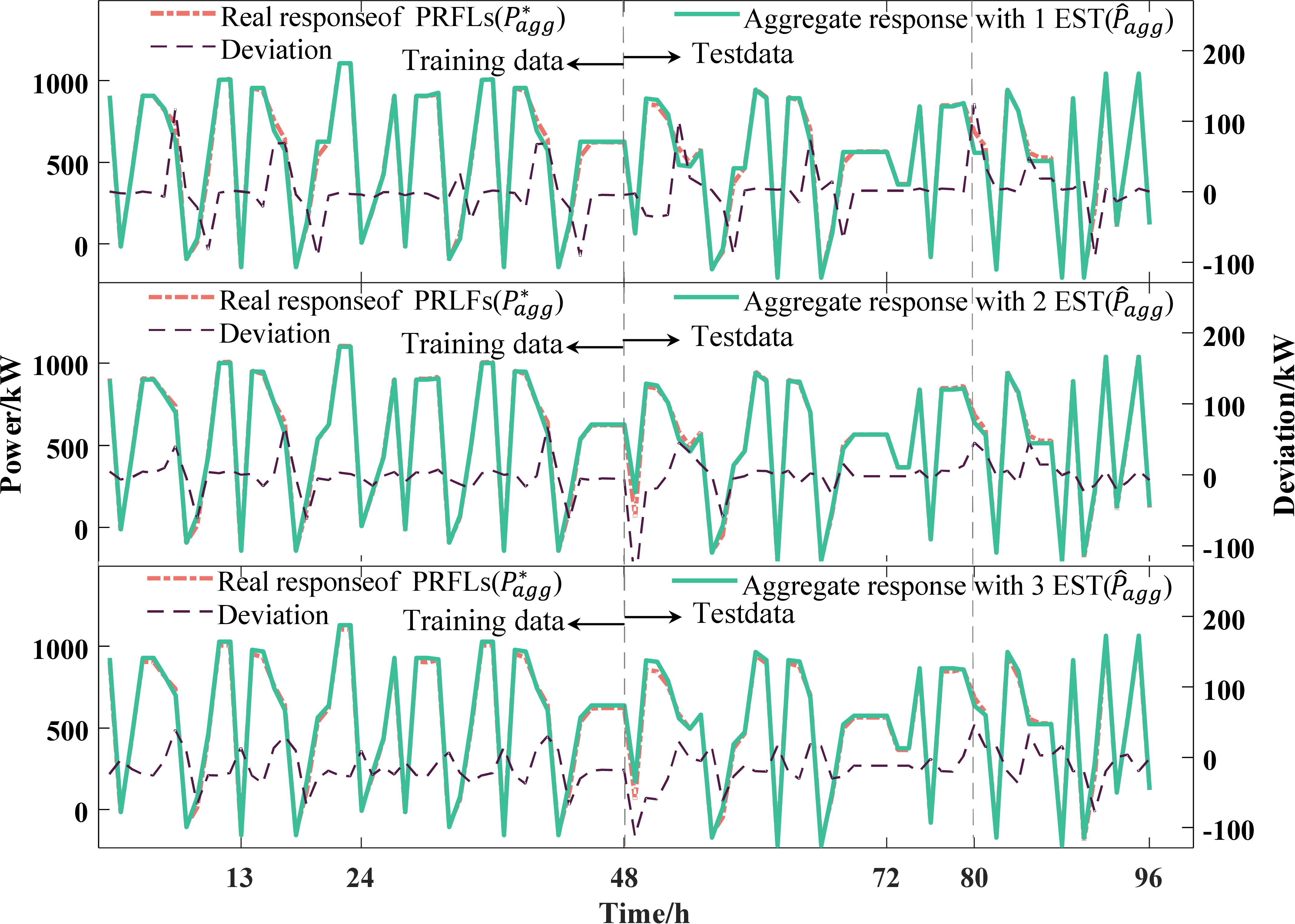}
\caption{Performance of  extended dynamic model with noise.}
\label{dynamic_ex_case}
\end{figure}

Given that PRFLs can use different response mechanism to actual price in reality, we examine the performance of the proposed theory and algorithm in dynamic case. Similar to the static response model, we evaluate the accuracy of dynamic and its extended model under different numbers of EST, displayed in Fig. \ref{dynamic_case} and Fig. \ref{dynamic_ex_case}. The error index NRMSE are demonstrated in Fig. \ref{NRMSE_static_dynamic}(b).

The dynamic model identification of AGM exhibits comparable accuracy to static model results, with precision improving proportionally to system flexibility within certain thresholds. Notably, our analysis reveals another meaningful phenomenon: dynamic aggregation outperforms the static one in the accuracy of identification. This disparity highlights the differential sensitivity of response mechanisms to AGM characteristics. Contrary to conventional expectations, increasing number of EST within surrogate AGM does not guarantee improved performance. This overfitting phenomenon likely originates from insufficient data diversity relative to model complexity. It further demonstrates that enhanced flexibility in surrogate AGM configurations proportionally increases parametric dimensionality, thereby imposing stricter requirements on data completeness. Model \eqref{static_response_model} and \eqref{dynamic_response_model} reveal that optimal responses are exclusively governed by the price direction vector \(\frac{\lambda^k}{\left|\lambda^k\right|}\)  rather than magnitude under fixed constraint parameters. Crucially, we conclude that comprehensive coverage of price direction permutations in reality ensures generalization error congruence between test and training set.


\vspace{-0.3cm}
\subsection{Comparison With Existing Algorithms}
\label{compare_test}
\begin{figure}[tb]
\centering
\footnotesize
\includegraphics[width=1\linewidth]{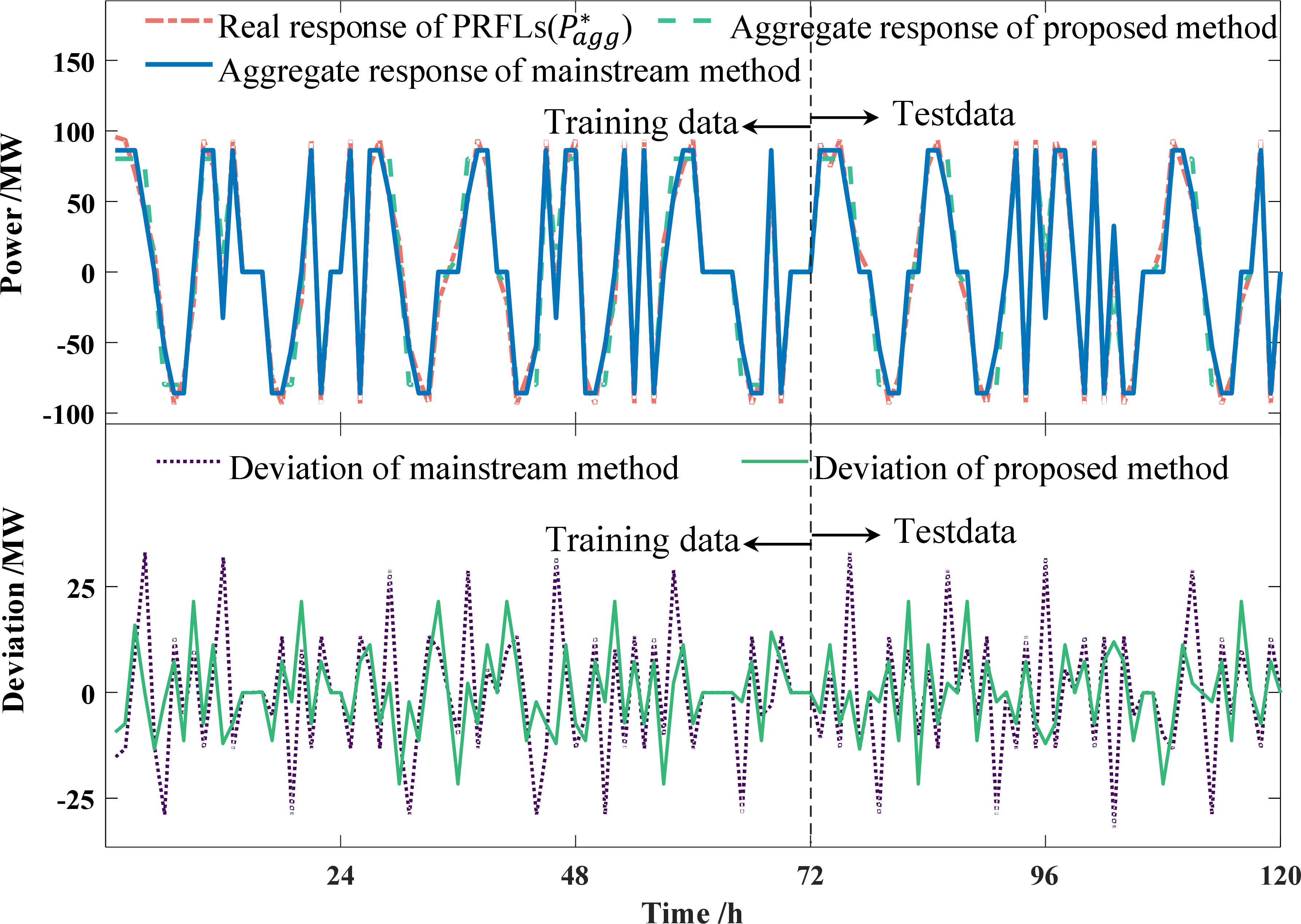}
\caption{Performance of the Newton-based method and the proposed method.}
\label{fig_compare}
\end{figure}
This comparative analysis evaluates a reduced-order static model focusing exclusively on boundary constraint identification for energy storage system parameters, neglecting storage efficiency and measurement noise. The simplified formulation permits solution via existing methods such as the Newton-based optimization – a mainstream approach demonstrating superior performance to machine learning benchmarks (multilayer perceptrons, support vector machines, LSTM networks). Fig. \ref{fig_compare} contrasts the proposed method against Newton-based optimization for single-component identification. The proposed method's deviations consistently reside within the Newton-based method's prediction envelope, demonstrating the enhanced aggregate response precision of  our method and also indicating the identification of superior parameters. The NRMSE of the Newton-based method in training data and test data is 0.0951 and 0.0940 relatively which is bigger than that of the proposed method 0.0619 in training data and 0.0633 in test data.

These numerical validations confirm the BayOpt based framework's efficacy for this inverse optimization problems, achieving 34.8\% error reduction over conventional methods. Furthermore, Gaussian process posterior enables to analyze the parameter identifiability and influence magnitude – a critical advancement for PRFL identification. 



\section{ Conclusion}
\label{section7}
This study aimed to identify the PRFLs with the operation data for PSOs. To achieve this target, this paper proposes a novel framework to identify parameters in AGM and integrates BayOpt with IO to fill the gap in current literature, which lacks both generalization ability and an effective algorithm for resolution. Moreover, this integration produces an identifiable posterior evaluation method that compensates for the shortcomings of prior methods. In general, we provides the promising theory and effective method for identification of PRFLs. Future work mentioned in this paper would be the optimal valuable price signal design and quantitative posterior evaluation of identifiability.


\appendix
\subsection{Proof of Theorem 1}
\label{proof1}

\begin{proof}
As mentioned before, the objective function in noise data can be calculated by:
\begin{align}
f_{nd}(\theta)&=\frac{1}{\left |\mathbf{N} \right |}  \sum_{n \in \mathbf{N}}\left \| {\Delta}^n+\mu_{agg}- \mu_{fix}\right \|_2^2\notag\\
&+\frac{2}{\left |\mathbf{N} \right |}  \sum_{n \in \mathbf{N}} (\overline{\Delta}^n)^\top(\varepsilon^n_{agg}- \varepsilon_{fix}^n-\mu_{agg}+\mu_{fix})\notag\\
&+\frac{1}{\left |\mathbf{N} \right |}  \sum_{n \in \mathbf{N}}(\left \|\varepsilon^n_{agg}- \varepsilon_{fix}^n\right \|_2^2 -\left \| \mu_{agg}- \mu_{fix}\right \|_2^2),\notag
\end{align}

To simplify the description, we denote the last two terms as $\xi_a$ and $\xi_b$ respectively. In practice, the observation of aggregate response power and the predicted fixed power are conducted as independent processes. Consequently, this two noise random variables associated with the two processes are independent:
\begin{align}
    &\mathbb{E}[(\varepsilon^n_{agg})^{\top}\varepsilon_{fix}^n]=\mathbb{E}(\varepsilon^n_{agg})^{\top}\mathbb{E}(\varepsilon_{fix}^n).\notag
\end{align}

For simplicity of description, we define variables $\xi_1^n$ and $\xi_2^n$:
\begin{align}
&\xi_1^n=( \overline{\Delta}^n)^\top(\varepsilon^n_{agg}-\varepsilon_{fix}^n-\mu_{agg}+\mu_{fix}) \notag\\
&\xi_2^n=\left \|\varepsilon^n_{agg}- \varepsilon_{fix}^n\right \|_2^2 -\left \| \mu_{agg}- \mu_{fix}\right \|_2^2 \notag
\end{align}

Further, the variables in two processes usually have the same mean value $\mu_{agg}=\mu_{fix}$ approximately equal to zero. Hence, we can calculate the mean value and variance of $\xi_1^n$ :
\begin{align}
&\mu_1=\mathbb{E}(\xi_1^n)=\mathbb{E}[(\overline{\Delta}^n)^\top(\varepsilon^n_{agg}-\varepsilon_{fix}^n-\mu_{agg}+\mu_{fix})] \notag\\
&=\mathbb{E}[(\overline{\Delta}^n)^\top(\varepsilon^n_{agg}-\varepsilon_{fix}^n)]=0 \notag.\\
&\sigma_1^2=( \overline{\Delta}^n)^\top(\Sigma_{agg}+\Sigma_{fix})\overline{\Delta}^n\notag.\\
&\xi_1^n \sim \mathcal{N}(0,\sigma_1^2)\notag
\end{align}

Similarly we calculate the mean and variance of $\xi_2^n$:
\begin{align}
&\mu_2=\mathbb{E}(\xi_2^n)=\mathbb{E}[\|(\varepsilon^n_{agg}- \mu_{agg})-(\varepsilon_{fix}^n-\mu_{fix})\notag\\
&+(\mu_{agg}- \mu_{fix})\|_2^2
-\left \| \mu_{agg}- \mu_{fix}\right \|_2^2]\notag
\end{align}

It is worth noting that $\mathbb{E}(\varepsilon^n_{agg}- \mu_{agg})=0$,$\mathbb{E}(\varepsilon_{fix}^n- \mu_{fix})=0$ and they are independent. Therefore, we can further calculate $\mu_2$ as:
\begin{align}
&\mu_2=\left \| \mu_{agg}- \mu_{fix}\right \|_2^2+\mathbb{E}[\|\varepsilon^n_{agg}- \mu_{agg}\|_2^2+ \|\varepsilon_{fix}^n-\mu_{fix}\|_2^2 \notag\\
&- 2*(\varepsilon^n_{agg}-\mu_{agg})^{\top}(\varepsilon_{fix}^n-\mu_{fix})]-\left \| \mu_{agg}- \mu_{fix}\right \|_2^2\notag\\
&=\left \| \mu_{agg}- \mu_{fix}\right \|_2^2+\mathrm{tr}(\Sigma_{agg})+\mathrm{tr}(\Sigma_{fix})-\left \| \mu_{agg}- \mu_{fix}\right \|_2^2\notag\\
&=\mathrm{tr}(\Sigma_{agg})+\mathrm{tr}(\Sigma_{fix}).\notag
\end{align}

For the $\xi_2^n$'s variance $\mathbb{V}(\xi_2^n)$, we can deduce its upper bound by the Cauchy inequality as:
\begin{align}
&\sigma_2^2=\mathbb{V}(\xi_2^n)=\mathbb{V}(\left \|\varepsilon^n_{agg}- \varepsilon_{fix}^n\right \|_2^2)\notag\\
&=\mathbb{E}(\left \|\varepsilon^n_{agg}- \varepsilon_{fix}^n\right \|_2^4)-[\mathbb{E}(\left \|\varepsilon^n_{agg}- \varepsilon_{fix}^n\right \|_2^2)]^2\notag\\
&\leq \mathbb{E}(\left \|\varepsilon^n_{agg}- \varepsilon_{fix}^n\right \|_2^4)\notag\\
&\leq T\mathbb{E}( \sum_{t \in \mathbf{T}}([\varepsilon^{n}_{agg}]_t- [\varepsilon_{fix}^{n}]_t)^4)\notag\\
&=T \sum_{t \in \mathbf{T}}(3\sigma_3^4+6\sigma_3^2\mu_3^2+\mu_3^4\notag)=M.\\
&\mu_3=\mathbb{E}([\varepsilon^{n}_{agg}]_t- [\varepsilon_{fix}^{n}]_t)=\mu_{agg}-\mu_{fix}\notag\\
&\sigma_3^2=\mathbb{V}([\varepsilon^{n}_{agg}]_t- [\varepsilon_{fix}^{n}]_t)=(\sigma_{agg})^2+(\sigma_{fix})^2.\notag
\end{align}

Wherein the $\mathrm{tr}(\Sigma)$ denotes the trace of the covariance matrix. Noting that the last two terms $\xi_a$ and $\xi_b$ can be expressed by the $\xi_1^n$ and $\xi_2^n$:
\begin{align}
\xi_a=\frac{2}{\left |\mathbf{N} \right |}  \sum_{n \in \mathbf{N}}\xi_1^n,\ 
\xi_b=\frac{1}{\left |\mathbf{N} \right |}  \sum_{n \in \mathbf{N}}\xi_2^n .\notag
\end{align}

From the law of large numbers, the $\xi_a$ and $\xi_b$ follow the normal distribution relatively:
\begin{align}
\xi_a\sim\mathcal{N}(0,\frac{4\sigma_1^2}{\left |\mathbf{N} \right |}),\ 
\xi_b\sim\mathcal{N}(0,\frac{\sigma_2^2}{\left |\mathbf{N} \right |}).\notag
\end{align}

Due to expectation of $\xi_1^n$ and $\xi_2^n$ exists and variance is bounded, we have:
\begin{align}
&\lim_{\left |\mathbf{N} \right | \to +\infty}\frac{4\sigma_1^2}{\left |\mathbf{N} \right |}=0
\notag,\\
&\lim_{\left |\mathbf{N} \right | \to +\infty}\frac{\sigma_2^2}{\left |\mathbf{N} \right |}\leq \lim_{\left |\mathbf{N} \right | \to +\infty}\frac{M}{\left |\mathbf{N} \right |}=0.\notag
\end{align}

Hence, from Chebyshev's inequality for $\forall\varepsilon>0$ we have:
\begin{align}
    \mathbb{P}(|f_{nd}(\hat\theta)-\frac{1}{|\mathbf{N}|}  \sum_{n \in \mathbf{N}}\left \| {\Delta}^n+\mu_{agg}- \mu_{fix}\right \|_2^2 \\\notag
    -(\mathrm{tr}(\Sigma_{agg})+\mathrm{tr}(\Sigma_{fix}))|>\varepsilon)=0.\notag
\end{align}

Finally, we draw an important conclusion:

\begin{equation}
\begin{aligned}
    &f_{nd}(\hat\theta)-\frac{1}{\left |\mathbf{N} \right |}  \sum_{n \in \mathbf{N}}\left \| {\Delta}^n+\mu_{agg}- \mu_{fix}\right \|_2^2\\
    &\stackrel{P}{\longrightarrow}\mathrm{tr}(\Sigma_{agg})+\mathrm{tr}(\Sigma_{fix}).\notag
\end{aligned}
\label{final_noise_obj}
\end{equation}

When the mean value of those noise is zero,i.e. $\mu_{agg}=0,\mu_{fix}=0$, the conclusion can be rewritten as:
\begin{equation}
\begin{aligned}
    (f_{nd}(\hat\theta)-f_{nf}(\hat\theta))_{\mathbf{N}}\stackrel{P}{\longrightarrow}\mathrm{tr}(\Sigma_{agg})+\mathrm{tr}(\Sigma_{fix}).\notag
\end{aligned}
\label{zeromean_obj}
\end{equation}

Comparing this objective function \eqref{final_noise_obj} with that \eqref{obj_nf} without noise, we can conclude the random variable $\varepsilon_{agg}$ and $\varepsilon_{fix}$ will not affect the solution of the model:
\begin{equation}
\begin{aligned}
    \lim_{\left |\mathbf{N} \right | \to +\infty}\hat{\theta}_{nd}=\lim_{\left |\mathbf{N} \right | \to +\infty}\hat{\theta}_{nf}.\notag
\end{aligned}
\end{equation}
\end{proof}
\subsection{Proof of Theorem 2}
\label{proof2}
\begin{proof}
To decompose the incremental components of $L_{n+1}$, we partition the Cholesky factor and its inverse as:
\begin{subequations}
\begin{equation}
    {{L_{n + 1}} = \left[ {\begin{array}{*{20}{c}}
{{L_n}}&0\\\notag
{{L_{21}}}&{{L_{22}}}
\end{array}} \right],
L_{n + 1}^{ - 1} = \left[ {\begin{array}{*{20}{c}}
{L_1^{ - 1}}&0\\\notag
{{L_3}^{ - 1}}&{L_2^{ - 1}},
\end{array}} \right]},
\end{equation}
\begin{equation}
K_{n + 1} = \left[ {\begin{array}{*{20}{c}}
{K_n}&{K_{n+1,1}}\\\notag
{K_{n+1,1}}^{\top}&{K_{n+1,n+1}}\notag
\end{array}} \right].
\end{equation}
wherein the submatrices derive from:
\begin{align}
&{L_{21}}= K_{n + 1,1}^ \top L_n^{ - \top},\ 
{L_{22}}= \sqrt {{K_{n + 1,n + 1}} - {L_{21}}L_{21}^ \top } ,\notag\\
 &K_{n + 1,1}=[k(\theta_1,\theta_{n+1}),k(\theta_2,\theta_{n+1}),\cdots,k(\theta_n,\theta_{n+1})]^{\top} ,\notag\\
 &K_{n + 1,n+1}=k(\theta_{n+1},\theta_{n+1}).\notag
\end{align}
Subsequently, the inverse components follow:
\begin{align}
    L_1^{ - 1} = L_n^{ - 1}, \ 
    L_2^{ - 1} = \frac{1}{L_{22}}, \ 
    L_3^{ - 1} =  - L_{22}^{ - 1}K_{n + 1,1}^ \top {({L_n}L_n^ \top )}^{ - 1}.\notag
\end{align}
Substituting into the Gaussian process update yields:
\begin{align}
K_{n + 1} = K_{n + 1,n + 1} - (L_{n + 1}^{ - 1}{K_{n + 1,1}})^TL_{n + 1}^{ - 1}K_{n + 1,1}.\notag
\end{align}
\end{subequations}
\end{proof}

\newcommand{\BIBdecl}{\setlength{\itemsep}{0.01 em}}
\bibliographystyle{IEEEtran}
\bibliography{IEEEabrv,reference.bib}

\end{document}